\documentclass[useAMS,usenatbib]{mn2e}
\usepackage{graphicx}
\usepackage{upgreek}
\usepackage{threeparttable}
\usepackage{siunitx}
\usepackage{url}

\def\SNR{\mbox{{SNR~J0509--6731}}}

\newcommand{\Halpha}{H${\alpha}$}

\newcommand{\sigmaD}{$\Sigma - D$}

\title[Radio--continuum study of LMC SNR J0509--6731]
  {Radio--continuum study of Large Magellanic Cloud Supernova Remnant J0509--6731}
\author[L. M. Bozzetto et al.]
  {L. M.~Bozzetto,$^1$ M. D.~Filipovi\'c,$^1$ D. Uro{\v s}evi{\'c},$^{2,3}$ R. Kothes,$^4$ \& E.~J.~Crawford$^1$\\
  $^1$School of Computing and Mathematics, University of Western Sydney
      \break Locked Bag 1797, Penrith South DC, NSW 1797, Australia\\
  $^2$Department of Astronomy, Faculty of Mathematics, University of Belgrade, Studentski trg 16, 11000 Belgrade, Serbia\\
  $^3$Isaac Newton Institute of Chile, Yugoslavia Branch \\
  $^4$National Research Council Canada, Herzberg Institute of Astrophysics, Dominion Radio Astrophysical Observatory, P.O. Box 248,\\
   ~Penticton, British Columbia V2A 6J9, Canada  }
\date{Released 2011 Xxxxx XX}

\pagerange{\pageref{firstpage}--\pageref{lastpage}} \pubyear{2002}

\def\LaTeX{L\kern-.36em\raise.3ex\hbox{a}\kern-.15em
    T\kern-.1667em\lower.7ex\hbox{E}\kern-.125emX}

\begin{document}

\label{firstpage}

\maketitle

\begin{abstract}
We present a detailed study of Australia Telescope Compact Array (ATCA) observations ($\lambda$ = 20, 13, 6 \& 3~cm) of supernova remnant (SNR) J0509--6731 in the Large Magellanic Cloud (LMC). The remnant has a ring morphology with brightened regions towards the south-western limb. We also find a second brightened inner ring which is only seen in the radio-continuum. The SNR is almost circular, with a diameter ranging from 7 to 8~pc, and a steep radio spectral index between 36 and 3~cm of $\alpha=-0.73\pm0.02$, which is characteristic of younger SNRs. We also report detection of radially orientated polarisation across the remnant at 6~cm, with a mean fractional polarisation level of  $P\cong$~(26~$\pm$~13)\%. We find the magnetic field ($\sim$168~$\mu$G) and \sigmaD\ ($\Sigma = $ $1.1\times 10^{-19}$~W m$^{-2}$~Hz$^{-1}$~sr$^{-1}$ , $D=$ 7.35~pc) to be consistent with other young remnants.\end{abstract}


\section{Introduction}

\begin{table*}
\begin{minipage}{165mm}
\caption{Summary of ATCA observations reduced and used in this study.}
\begin{threeparttable}
\label{tbl-obs}
\begin{tabular}{@{}cccccccccc}
\hline
Date & Scan time & Right Ascension & Declination & Array & Frequencies & BWidth & Chan & Project\\
& (minutes) & & & & (MHz) & (MHz) & & \\
\hline
2011-Nov-16 	& 49.7 	& 5h  9m 31.00s & -67\degr31\arcmin16.20\arcsec & EW367 	& 5,500, 9,000 & 2048.0 	& 2049 	& C634$^a$\\
2011-Nov-15 	& 15.0	& 5h  9m 31.00s & -67\degr31\arcmin16.20\arcsec & EW367 	& 5,500, 9,000 & 2048.0 	& 2049 	& C634$^a$\\
2010-Nov-29 	& 98.0 	& 5h  9m 30.00s & -67\degr30\arcmin60.00\arcsec & 6A 		& 5,500, 9,000 & 2048.0 	& 2049	& C2367\\
2010-Nov-28 	& 50.3 	& 5h  9m 30.00s & -67\degr30\arcmin60.00\arcsec & 6A 		& 5,500, 9,000 & 2048.0 	& 2049	& C2367\\
2005-Jun-24 	& 819.7 	& 5h  9m 51.48s & -67\degr16\arcmin22.17\arcsec & 6B 		& 1,384, 1,472 & 128.0 	& 33 		& C1395\\
2005-Apr-18 	& 819.8 	& 5h  9m 51.48s & -67\degr16\arcmin22.17\arcsec & 1.5A 	& 1,384, 1,472 & 128.0 	& 33 		& C1395\\
1997-Aug-11	& 643.7 	& 5h  9m 30.00s & -67\degr30\arcmin60.00\arcsec & 750B 	& 4,800, 4,928 & 128.0	& 33	 	& C479\\
1994-Sep-23 	& 57.0 	& 5h  9m 31.00s & -67\degr31\arcmin15.00\arcsec & 1.5D 	& 1,380, 2,378 & 128.0 	& 33 		& C354\\
1994-Sep-22 	& 482.3 	& 5h  9m 31.00s & -67\degr31\arcmin15.00\arcsec & 1.5D 	& 1,380, 2,378 & 128.0 	& 33 		& C354\\
1994-Sep-17 	& 260.7 	& 5h  9m 31.00s & -67\degr31\arcmin15.00\arcsec & 1.5B 	& 1,380, 2,378 & 128.0 	& 33 		& C354\\
\hline
\end{tabular}
\begin{tablenotes}
\item $^a$ -- The observing procedure in this project is described in the text.
\end{tablenotes}
\end{threeparttable}
\end{minipage}
\end{table*}

Supernova remnants (SNRs) play a vital role in the universe, enriching the interstellar medium (ISM) and significantly influences the ISMs evolution, structure and physical properties. The study of SNRs in our own Galaxy is not ideal due to difficulties in estimating accurate distances (which inhibits accurate analysis such as extent and surface brightness) and the high level of absorption in the direction of the Galactic plane. As an alternative, the Large Magellanic Cloud (LMC) at a proximity of 50~kpc \citep{2006ApJ...652.1133M} is a near ideal galaxy for the study of SNRs due to its high active star forming regions (such as 30 Dor) and location outside of the Galactic plane at an angle of 35$^\circ$ \citep{2001AJ....122.1807V}. Its distance from earth allows us to assume that objects located within are at approximately the same distance, aiding in various analysis methodologies.\\

In the radio-continuum, SNRs predominately emit non-thermal continuum emission and generally exhibit a spectrum of $\alpha\sim-0.5$ (defined by $S\propto\nu^\alpha$). Although this can vary as there exists a wide variety of SNRs in different stages of evolution and expanding in different environments \citep{1998A&AS..127..119F}. \\

In this paper, we present new radio-continuum observations of \SNR, along with archival radio-continuum, X-ray \& optical observations. This source was originally classified by \citet{1981ApJ...248..925L} as an SNR in their X-ray survey using the Einstein observatory, recording a position of {RA~(B1950)=05$^h$09$^m$28$^s$} and \mbox{DEC~(B1950)=--67\degr34\arcmin55\arcsec}. \citet{1982ApJ...261..473T}  estimate a X-ray size of $\sim$25\arcsec, an optical size of 25\arcsec\ and estimated a shock velocity of $>$ 3600~km~s$^{-1}$. No object was found at this position in the 408~MHz image by Clark, Little \& Mills (1976). However, reanalysis of the 408~MHz survey data by \citet{1982ApJ...261..473T}, found weak emission at this point finding a flux density measurement of 95$\pm$15~mJy. \citet{1982ApJ...261..473T} also observed this object at 5~GHz and measured a flux density of 30$\pm$3~mJy. They note that on a surface brightness to diameter (\sigmaD) diagram, \SNR\ fell below the mean line by a factor of 13, which is comparable to young Galactic SNRs. They comment that this low \sigmaD\ might be a result of differences in the electron acceleration process in Balmer dominated remnants. The remnant is described as a Balmer dominated SNR expanding into a region with a relatively low density ({\it n}$_H$ $\leq$ 0.02$^{-3}$) of neutral hydrogen and argue for a Type Ia supernova (SN). \citet{1983ApJS...51..345M} measure an X-ray size of 27\arcsec\ and a radio spectrum of $\alpha$ = --0.46.  \citet{1984ApJ...281..593F} estimates a shock temperature of 3.1~KeV, an age of 900 yr, total swept up mass of 26~{\it M}$_{\sun}$ and a shock velocity of 1600~km~s$^{-1}$, which is well below that proposed by \citet{1982ApJ...261..473T} of $>$3600.  \citet{1984AuJPh..37..321M} record a 843~MHz flux density of 82~mJy, updating the spectral index to --0.48 and also record a surface brightness of $>$6.4$\times$10$^{20}$ ~W m$^{-2}$~Hz$^{-1}$~sr$^{-1}$. \citet{1988ApJ...327..156V} also argues for a younger remnant, commenting that the small diameter indicates an age of $\leq$1000~yr.  \citet{1988AJ.....96.1874C} give this SNR OB association 400~pc to LH38 and class this remnant as population II.  \citet{1991ApJ...375..652S} records a shock velocity $>$2000~km~s$^{-1}$ and is in agreement with an age of $\leq$1000 yr.  \citet{1995ApJ...444L..81H} notes that there is strong emission of elements from silicon to neon and argues for Type Ia SN.  \citet{1999A&AS..139..277H} record an extent of 9.1\arcsec\ and give this SNR the association [HP99] 542.  \citet{2004ApJ...608..261W} confirmed that the SN ejecta had an abundance distribution consistent with Type Ia SN explosion models. They also found that the reverse shock is propagating back into the Fe-rich ejecta and suggests that the brighting in the southwest is due to enhanced density in or a deeper penetration of the reverse shock into the into a portion of the ejecta shell and may be caused enhanced ambient density or intrinsic asymmetry in the explosion itself. \citet{2005Natur.438.1132R} confirmed the Type Ia classification using light echo spectra and also established it as a SN1991T-type energetic event. Additionally, light echo apparent motion was used to estimate the age of the SNR to be 400$\pm$120 yr. \citet{2005MNRAS.360...76A} used a 1~GHz flux density of 70~mJy  to estimate a surface brightness--diameter of (\sigmaD) = ($4.2\times 10^{-20}$~W m$^{-2}$~Hz$^{-1}$~sr$^{-1}$, 7~pc). \citet{2007ApJ...664..304G} estimate an age of 295 - 585 yr, a shock velocity of V$_s$ $\geq$4000~km~s$^{-1}$, they detect broad Ly $\beta$ emission and classify this object as a non-radiative (adiabatic) of Type Ia. \citet{2008ApJ...680.1149B} found an age of $\sim$400 yr, kinetic energy of 1.4$\times$10$^{51}$ ergs and concluded that the X-ray properties of \SNR\ were consistent with models of an energetic 91T-type SN Ia explosion. \citet{2008PASJ...60S.453S} states \SNR\ is thought to be dominated by thermal dust continuum  with T(dust) 94$\pm$3~K and a Dust mass of 8.7$\pm$2.5$\times$10$^{-5}$ solar masses. \citet{2008A&A...490..223K} also find that the reverse shock has recently reach the iron layers of the ejecta and are in agreement with previous studies regarding the brightening in the southwest resulting from an asymmetric explosion or density enhancement in the ISM. Models in this study were in good agreement with the observations with circumstellar density of 3$\times$10$^{-25}$ g/cm$^3$, age of $\sim$400 yr and velocity of $\sim$5000 km s$^{-1}$. \citet{2010AJ....140..584D} found no association between this remnant and a YSO, nor the molecular clouds. \citet{2012Natur.481..164S} found no ex-companion star to a visual magnitude limit of 26.9 within a radius of 1.43\arcsec, which they state would infer a double degenerate SN system. \citet{2012ApJ...759...56D} and \citet{2012ApJ...758..123W} discuss the possibility of this SNR still being the result of a single degenerate explosion.

The observations, data reduction and imaging techniques are described in Section~2. The astrophysical interpretation of newly obtained moderate-resolution total intensity and polarimetric images in combination with archival Chandra X-ray and HST \Halpha\ observations are discussed in Section~3.

\section{Observations and data reduction}

Five Australia Telescope Compact Array (ATCA) projects (C1395, C354, C479, C634 and C2367; at wavelengths of  20~cm, 20/13~cm, 6~cm, 6/3~cm and 6/3~cm respectively) were reduced and analysed in this study. A summary of these projects can be seen in Table \ref{tbl-obs}.  Project C634 contain our observations of this SNR, which were taken on the 15$^\mathrm{th}$ and 16$^\mathrm{th}$ of November 2011. These observations were taken by the ATCA using the CABB receiver with the array configuration EW367, at wavelengths of 3 and 6~cm ($\nu$=9000 and 5500~MHz). The observations were carried out in the so called ``snap-shot'' mode, totalling $\sim$50 minutes of integration over a 14 hour period. Source PKS~B1934-638 was used for primary (flux density) calibration and source PKS~B0530-727 was used for secondary (phase) calibration. At 6~cm, the shorter baselines from the EW367 observations were complemented by observations taken from project C2367, which uses a longer baseline array configuration (6A; Table \ref{tbl-obs}), allowing for a higher resolution image. However, we were unable to make use of the 3~cm data from ATCA project C2367 due to strong interference. This lack of data meant we lost the longer baselines and as a result, no high resolution image is available at this wavelength. 

The \textsc{miriad}\footnote{http://www.atnf.csiro.au/computing/software/miriad/} \citep{1995ASPC...77..433S} and \textsc{karma}\footnote{http://www.atnf.csiro.au/computing/software/karma/} \citep{1995ASPC...77..144G} software packages were used for reduction and analysis. More information on the observing procedure and other sources observed in this project can be found in \citet{2012MNRAS.420.2588B, 2012RMxAA..48...41B, 2012SerAJ.185...25B, 2013MNRAS.432.2177B} and \citet{2012A&A...540A..25D}.

Images were formed using \textsc{miriad} multi-frequency synthesis \citep{1994A&AS..108..585S} and natural weighting. They were deconvolved with primary beam correction applied. The same procedure was used for both U and Q stokes parameter maps.

We measured the flux density of \SNR\ from 11 separate images between 36~cm and 3~cm, which are summarised in Table \ref{tbl-flux}. We obtain five of these flux density measurements from available mosaics; at 36~cm from the Molonglo Synthesis Telescope (MOST) mosaic image (as described in \citealt{1984AuJPh..37..321M}) and from the SUMMS mosaic image \citep{2008yCat.8081....0M}, 20~cm from the mosaic by \citet{2007MNRAS.382..543H}. We also used 6~cm and 3~cm mosaics published by \citet{2010AJ....140.1567D}. The remaining six measurements were taken from the data reduced and analysed in this study using the projects listed in Table \ref{tbl-obs}. Errors in flux density measurements predominately arose from uncertainties in defining the `edge' of the remnant. However, we estimate these errors to be $<$10\% (with the exception of the 73~cm measurement, where the associated error is given by \citealt{1982ApJ...261..473T}). Using the flux density measurements in Table \ref{tbl-flux} (73 -- 3~cm), we estimate a spectral index of $\alpha = -0.59$. However, it can be seen that the spectrum breaks at 73~cm, where the recorded flux density is at a level well below that which is expected (by $\sim$50\%). Low frequency absorption can result in this `break', either through synchrotron self absorption or thermal absorption. A low frequency turnover assumed to be from free-free absorption was found for 7 SNRs in M82 \citep{1997MNRAS.291..517W} at levels comparable to our 408~MHz turnover for \SNR. However, M82 has an environment significantly denser than the relatively rarified environment of \SNR, and therefore, the turnover is expected to occur at higher frequencies in this denser environment. The more probable explanation for this break is observational effects or an issue with the measurement. Omitting this outlying value from our calculation results in a steeper spectral index with a value of $\alpha = -0.73 \pm 0.02$.

\begin{table*}
\begin{minipage}{135mm}
\caption{Integrated flux densities of \SNR.}
\begin{threeparttable}
\label{tbl-flux}
\begin{tabular}{@{}cccccccl}
\hline
$\lambda$      & $\nu$ & ATCA & R.M.S  & Beam Size & S$_\mathrm{Total}$ & $\Delta$S$_\mathrm{Total}$ & Reference \\
(cm) & (MHz) & Project & (mJy) & (\arcsec) & (mJy) & (mJy) &   \\
\hline
73 & 408 & MOST & 40 & 157.3$\times$171.6 & 95 & 15 & \citet{1982ApJ...261..473T}\\
36 & 843 & MOST$^a$ & 0.4 & 46.4$\times$43.0 & 111 & 11 & This work\\
36 & 843 & SUMMS$^b$ & 1.5 & 48.5$\times$45.0 & 109 & 11 & This work\\
20 & 1373 & C354 & 0.3 & 21.2$\times$17.3 & 73 & 7 & This work\\
20 & 1377 & C373$^c$ & 0.7 & 40.0$\times$40.0 & 80 & 8 & This work\\
20 & 1381 & C1395 & 0.7 & 13.0$\times$12.2 & 79 & 8 & This work\\
13 & 2377 & C354 & 0.3 & 12.3$\times$10.1 & 51 & 5 & This work\\
6 & 4800 & Multiple$^d$ & 1.0 & 35.0$\times$35.0 & 30 & 3 & This work\\ 
6 & 4800 & C479 & 0.3 & 28.6$\times$11.8 & 30 & 3 & This work\\ 
6 & 5000 & Parkes &  --- & 300$\times$300 & 30 & 3 & \citet{1982ApJ...261..473T}\\
6 & 5500 & C634, C2367 & 0.1 & 2.6$\times$2.3 & 31 & 3 & This work\\ 
3 & 8640 & Multiple$^d$ & 1.0 & 22.0$\times$22.0 & 19 & 2 & This work\\
3 & 9000 & C634 & 0.2 & 22.7$\times$16.0 & 20 & 2 & This work\\
\hline
\end{tabular}
\begin{tablenotes}
\item $^a$ -- From the image described in \citet{1984AuJPh..37..321M}
\item $^b$ -- From the image described in \citet{2008yCat.8081....0M}
\item $^c$ -- From the image described in \citet{2007MNRAS.382..543H}
\item $^d$ -- From the image described in \citet{2010AJ....140.1567D}
\end{tablenotes}
\end{threeparttable}
\end{minipage}
\end{table*}

\section{Results and Discussion}

\begin{figure}
\centering\includegraphics[trim=0 20 0 0, angle=-90, scale=.348]{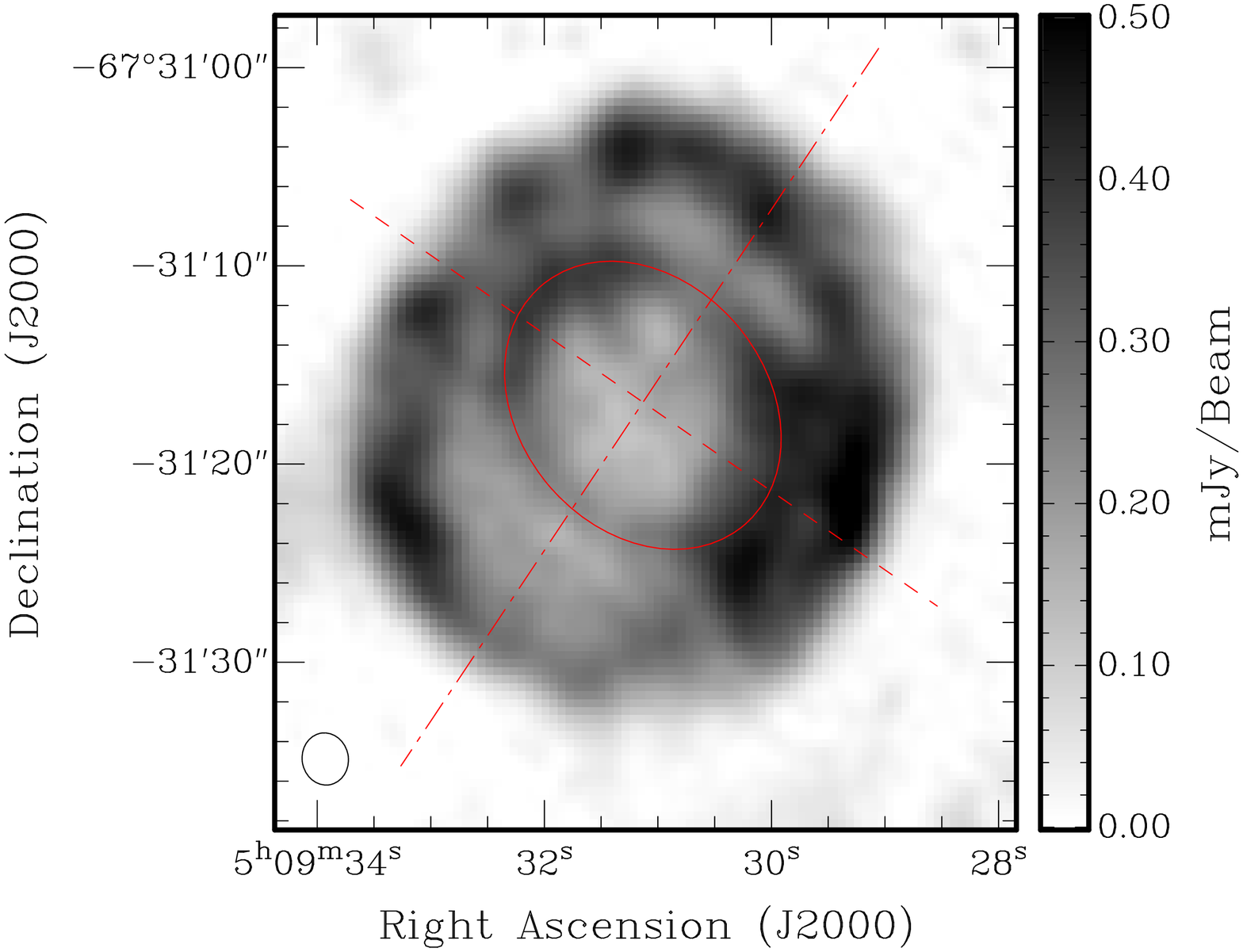}
\centering\includegraphics[scale=.25,angle=-90]{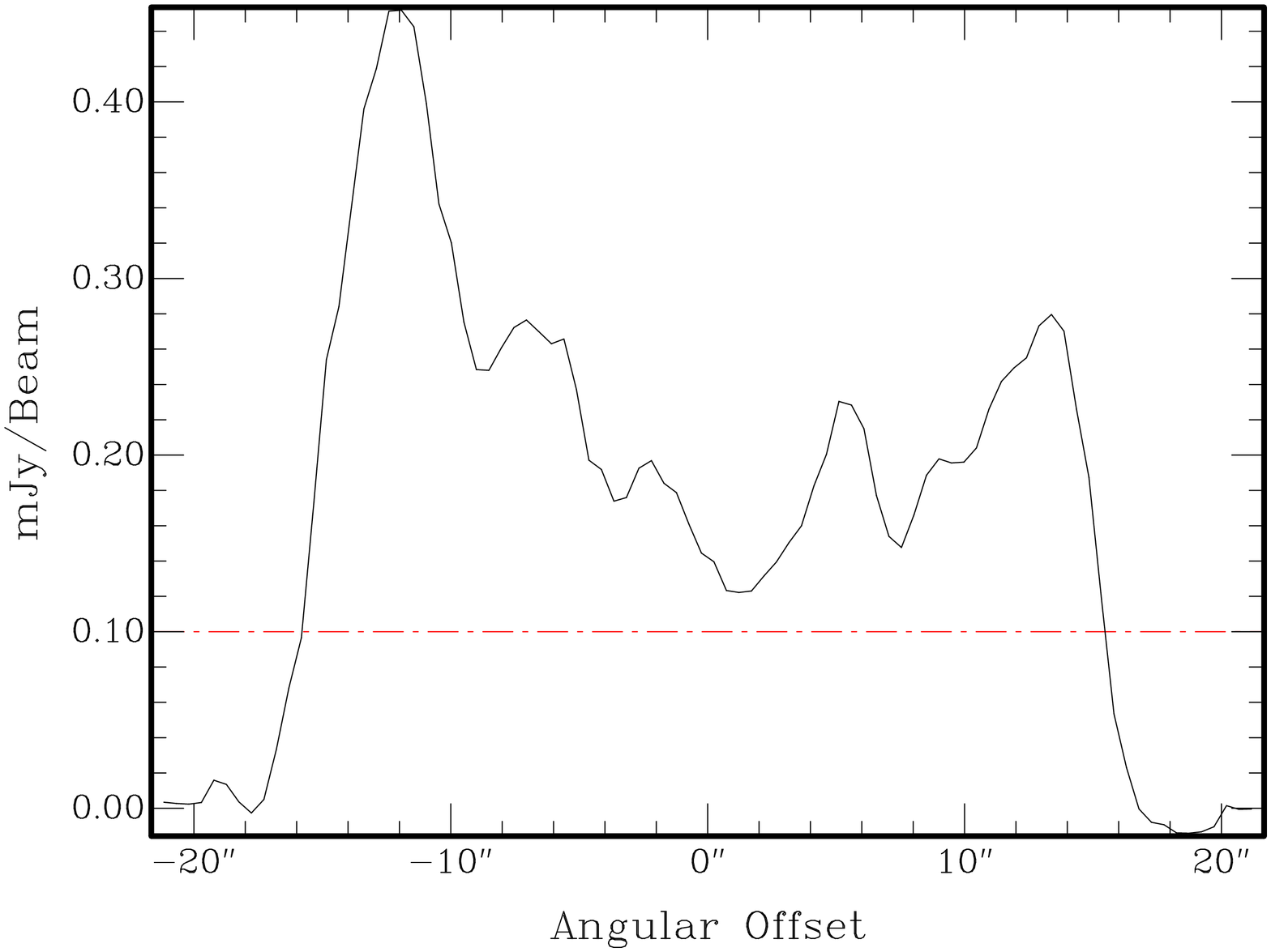}
\centering\includegraphics[scale=.25,angle=-90]{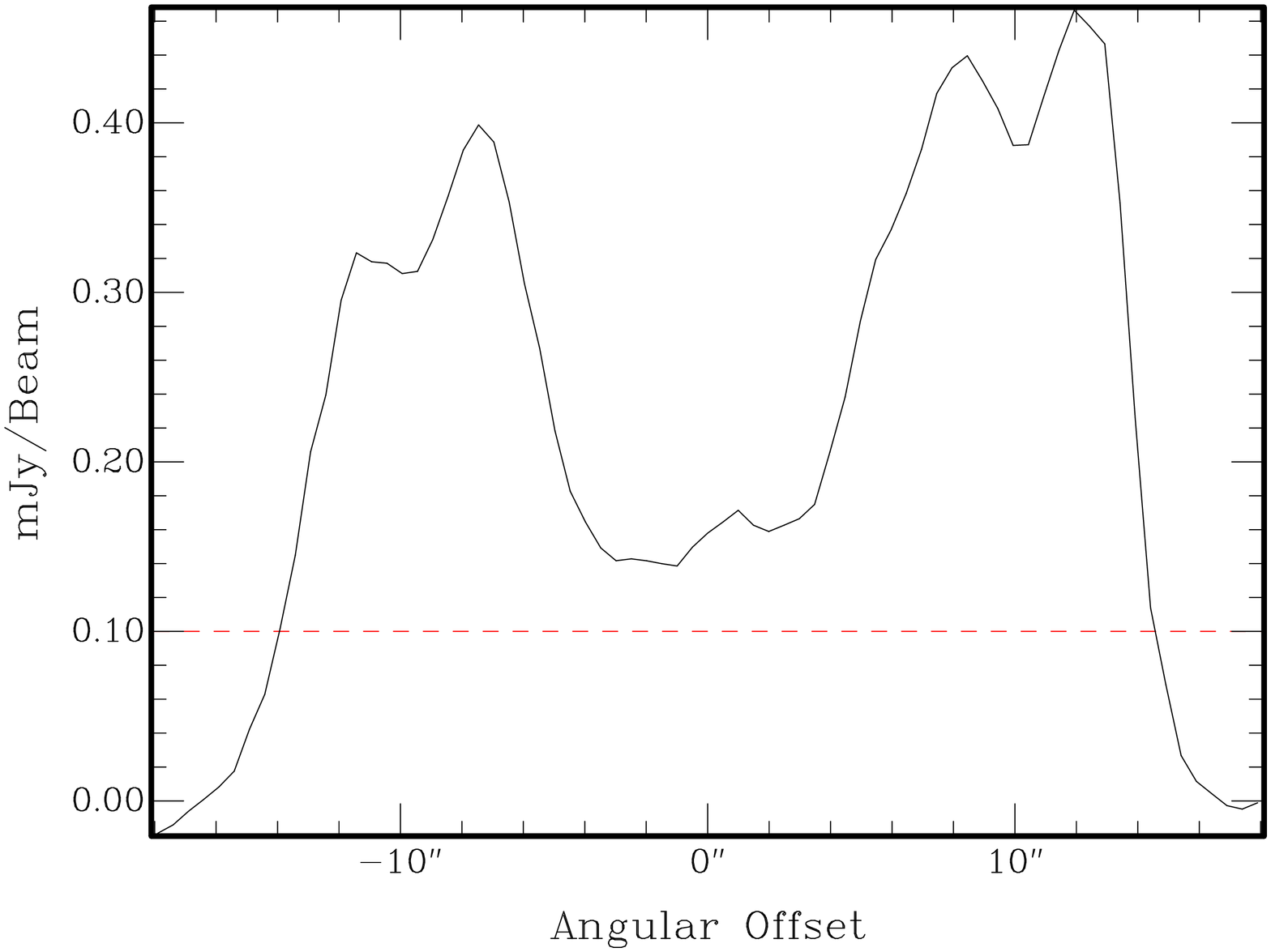}
\caption{The top image shows the 6~cm intensity image of \SNR\ overlaid with the approximate major (NW--SE) and minor (NE--SW) axis. The middle and lower images show the 1-dimensional cross-section along the overlaid lines in the top image, with a superimposed line at 3$\sigma$.  
\label{extent}}
\end{figure}

\SNR\ exhibits a ring-like morphology (Fig.~\ref{extent}), centred at \mbox{RA~(J2000)=05$^h$09$^m$31.0$^s$}, \mbox{DEC~(J2000)=--67\degr31\arcmin16.4\arcsec.} We estimate the spatial extent of \SNR\ (Fig.~\ref{extent}) at the 3$\sigma$ (Table~\ref{tbl-flux}; Col.~4) level (0.1~mJy) along the major (NW--SE) and minor (NE--SW) axes (PA=--34\degr). Its size at 6~cm (5500~MHz) is 31\arcsec$\times$29\arcsec$\pm$1\arcsec\ (8$\times$7~pc with 0.25~pc uncertainty in each direction). We estimate the ring thickness of \SNR\  to $\sim$6\arcsec\  at 6~cm, about 40\% of the SNR's radius. 

We find a centrally brightening ring in the interior of this remnant (Fig. \ref{extent}), something that is not common among SNRs. We estimate the size of this ring at 6~cm to be 16\arcsec$\times$12\arcsec$\pm$1\arcsec\ (4$\times$3~pc with 0.25~pc uncertainty in each direction) at PA=50\degr.

\begin{figure}
\centering\includegraphics[trim=0 0 0 0,scale=.5]{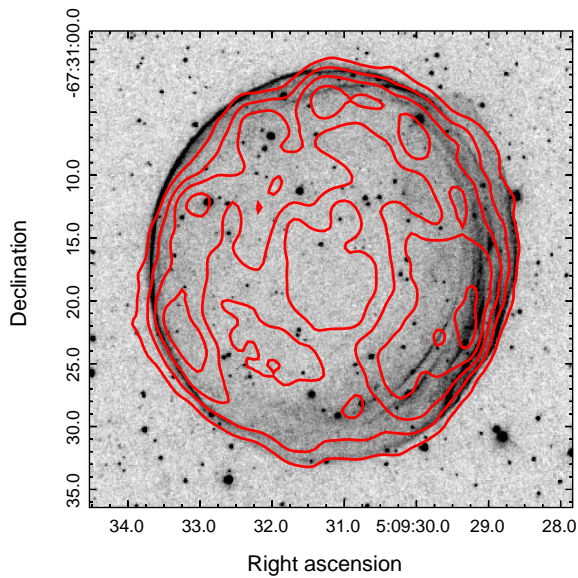}
\caption{HST H$\alpha$ image of \SNR\ overlaid with 6~cm ATCA contours. The contours are 3, 6, 9, 12 \& 15$\sigma$ (where $\sigma$ = 33~$\mu$Jy).
 \label{radVSha}}
\end{figure}

There is evident correlation between our 6~cm (5500~MHz) radio-continuum emission and the optical \Halpha\ emission (Hubble Space Telescope; PropID 11015) for this remnant (Fig.~\ref{radVSha}). This is particularly evident towards the south-western limb of the SNR (where radio emission is the strongest), where we can see the radio 3$\sigma$ contour closely following the edge of the optical \Halpha\ emission. The astrometry involved in aligning all images in this paper is within 1\arcsec.

\begin{figure}
\centering\includegraphics[trim=0 0 0 0, scale=.5]{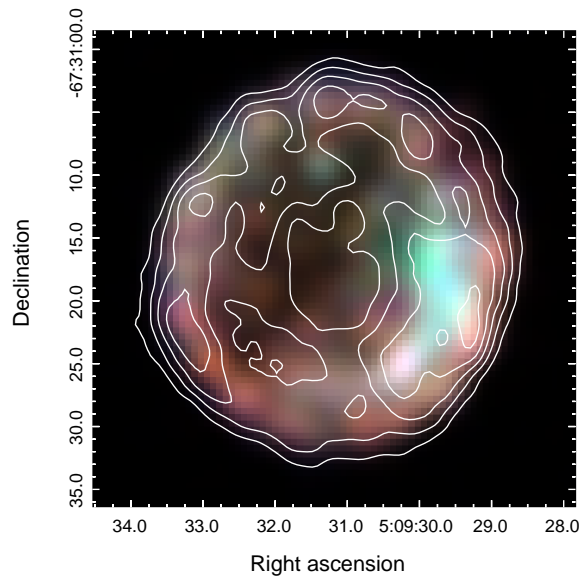}
\caption{ Chandra X-ray colour composite image of \SNR\ at energy levels 0.3--0.6 keV (red) 0.6--0.95 keV (green) and 0.95--7.0 keV (blue). The image has been smoothed using a gaussian filter ($\sigma$ = 2 pixel). ATCA radio contours (at 6~cm) have been overlaid at levels of 3, 6, 9, 12 \& 15$\sigma$ (where $\sigma$~=~33~$\mu$Jy).
 \label{radvsxray}}
\end{figure}

We also find similarities between our 6~cm (5500~MHz) radio-continuum emission and  0.3--7.0~keV X-ray emission (Chandra; observation ID [ObsID] 776\footnote{ Taken from \url{http://hea-www.harvard.edu/ChandraSNR/snrcat_lmc.html}}) as seen in Fig.~\ref{radvsxray}. 

The optical \Halpha\ emission shows highly compressed filaments, denoting at high angular resolution the location of the forward shock moving into the ISM, outlying the ellipsoidal shell region interior to which the smooth, low compressed radio and X-ray emission comes.

\begin{figure}
\centering\includegraphics[scale=.55]{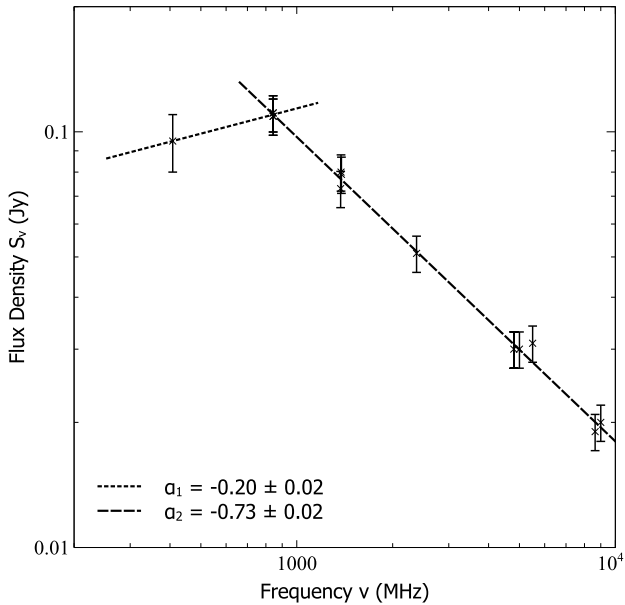}
\caption{Radio-continuum spectrum of \SNR. The markers represent error margins of 10\%.
 \label{spcidx}}
\end{figure}

The non-thermal nature of this remnant in the radio-continuum is confirmed in the spectral energy distribution (SED), shown in Fig. \ref{spcidx}, where $\alpha _2 = -0.73\pm0.02$. This value is steeper in comparison with typical values of $\alpha =-0.5$ for LMC SNRs \citep{1998A&AS..127..119F} and is more consistent with young SNRs, shown in Table \ref{tbl-altsnrs}. This is in agreement with current estimation of the remnants age, which places it at $\sim$400 yr \citep{2005Natur.438.1132R, 2007ApJ...664..304G, 2008A&A...490..223K, 2008ApJ...680.1149B}. 

\begin{figure}
\centering\includegraphics[angle=0,scale=.44,trim=0 0 0 0,clip]{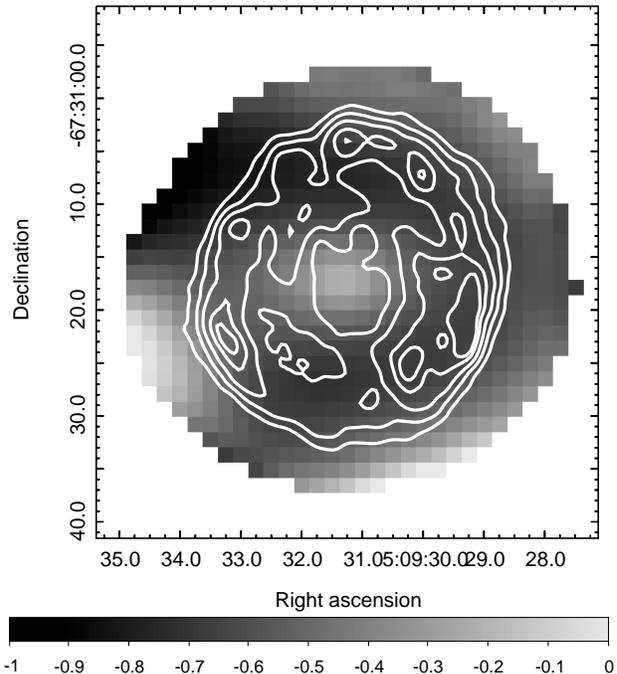}
\caption{Radio-continuum spectrum map of \SNR\ between 13~cm and 6~cm. The sidebar quantifies the spectral index scale. ATCA radio contours (at 6~cm) have been overlaid at 3, 6, 9, 12 \& 15$\sigma$ (where $\sigma$ = 33~$\mu$Jy).
 \label{spcmp}}
\end{figure}

A spectral index map was created between 13~cm and 6~cm (Fig. \ref{spcmp}) to show the spacial spectral variations in the remnant. This was achieved by convolving and re-gridding the 6~cm image with the tasks {\tt regrid} and {\tt convol}, to match the size and resolution of the 13~cm image, which had the poorest resolution and thus allowing no oversampling to occur. A spectral index map was then created using these maps from both observed frequencies. This was done using the \textsc{miriad} task {\tt maths}, which calculated the spectral index\footnote{spectral index $\alpha$ is defined by $S_{\nu}=\nu^{\alpha}$, where $S_{\nu}$ is the integrated flux density and $\nu$ is the frequency.}  ($\alpha$) of each pixel above a level of 3$\sigma$. Pixels below this level were blanked in the spectral index map. We note to two distinctive and opposite regions of somewhat steeper spectra ($\sim\alpha$=--0.7) marked in yellow around northern and southern regions of the SNR. 

\begin{figure}
\centering\includegraphics[angle=-90,scale=.38]{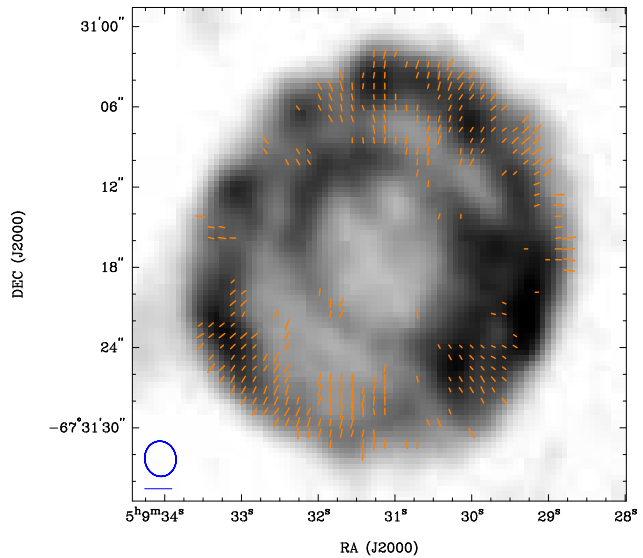}
\caption{B-field polarisation vectors overlaid on 6~cm ATCA image of \SNR. The blue ellipse in the lower left corner represents the synthesised beamwidth of 2.6\arcsec$\times$2.3\arcsec and the blue line below the ellipse represents a polarization vector of 100\%.
 \label{polar}}
\end{figure}

A fractional polarisation image was created at 6~cm using \textit{Q} and \textit{U} parameters (Fig.~\ref{polar}). A signal-to-noise cut-off of 2$\sigma$ was used for the \textit{Q} and \textit{U} images, while a level of 6$\sigma$ was used for the intensity image. Values that fall below these cut-off levels are blanked in the output image.  The length of the vectors have been reduced by 50\% and placed every 1.5 pixels for display purposes. The mean fractional polarisation was calculated using flux density and polarisation:\\

P=$\frac{\sqrt{S_{Q}^{2}+S_{U}^{2}}}{S_{I}}$\\

\noindent where $S_{Q}, S_{U}$ and $S_{I}$ are integrated intensities for the \textit{Q}, \textit{U} and \textit{I} Stokes parameters. We estimate a mean fractional polarisation value of  $P=$~26~$\pm$~13\% at 6~cm. The magnetic field of the remnant at 6~cm appears to be radially oriented, which is to be expected from Rayleigh-Taylor instabilities in the decelerating remnant (Gull 1975; Chevalier 1976). This is consistent with similarly young SNRs in our own Galaxy, as well as in the LMC (for e.g. those listed in Table \ref{tbl-altsnrs}).

\begin{table*}
\begin{minipage}{135mm}
\caption{Comparison of \SNR\ to similar remnants.}
\begin{threeparttable}
\label{tbl-altsnrs}
\begin{tabular}{@{}lccccl}
\hline
Name & Age & $\alpha$$^*$ & \textit{P} & \textit{P}$_\lambda$ & Reference \\
 & (yr)  & & (\%) & (cm) &   \\
\hline
0509--67.5 & $\sim$400 & $-0.73$ & 26$\pm$13 & 6 & This work\\
\hline
Cassiopeia A & --- & $-0.77$ & 8-10 & 6 & \citet{1995ApJ...441..300A}\\
Tycho & $\sim$441 &  $-0.65$ & 20-30$^a$ & 6 & \citet{1991AJ....101.2151D}\\
Kepler & $\sim$409 & $-0.64$ & 6 & 6 & \citet{2002ApJ...580..914D}\\
SN 1006 & $\sim$1000 & $-0.6$ & 17$^d$ & 20 & \citet{2013AJ....145..104R}\\
\hline
N132D & $\sim$2500$^b$ & $-0.70$ & 4 & 6 & \citet{1995AJ....109..200D}\\
0519-6902 & $\sim$600$^c$ & $-0.53$ & 2 & 6 & \citet{2012SerAJ.185...25B}\\
\hline
\end{tabular}
\begin{tablenotes}
\item $^*$ -- Galactic spectral indices came from the catalogue by \citet{2009BASI...37...45G}.
\item $^a$ -- Based on the mean polarisation found for the brightened limbs
\item $^b$ -- \citet{2011Ap&SS.331..521V}
\item $^c$ -- \citet{2006ApJ...642L.141B}
\item $^d$ -- Higher polarisation (near the theoretical limit of $\sim$70\%) was found in regions of weaker radio emission.
\end{tablenotes}
\end{threeparttable}
\end{minipage}
\end{table*}

Without reliable polarisation measurements at a second frequency we cannot determine the Faraday rotation and thus cannot deduce the magnetic field strength. However, we make use of the equipartition formula as given by \citet{2012ApJ...746...79A} to estimate the magnetic field strength of this SNR. This formula is based on the \citet{1978MNRAS.182..443B} diffuse shock acceleration (DSA) theory. This derivation is purely analytical, accommodated especially for the estimation of magnetic field strength in SNRs. The average equipartition field over the whole shell of \SNR\ is $\sim$168~$\mu$G with an estimated minimum energy of E$_{min}$ = 1.2$\times$10$^{49}$ ergs (see \citet{2012ApJ...746...79A}; and corresponding ``calculator"\footnote{The calculator is available at http://poincare.matf.bg.ac.rs/\~{}arbo/eqp/ }). This value is typical of young SNRs with a strongly amplified magnetic field.\\

The position of \SNR\ at the surface brightness to diameter (\sigmaD) diagram ($\Sigma$ = $1.1\times 10^{-19}$~W m$^{-2}$~Hz$^{-1}$~sr$^{-1}$, $D$ = 7.35~pc) by \citet{2004A&A...427..525B}, suggests that this remnant is in the transition phase between late free expansion and early Sedov phase, with an explosion energy of $\sim$0.25$\times$10$^{51}$, which evolves in an environment with a density of $\sim$0.3~cm$^{-3}$. This estimate of minimum explosion energy is lower than that found by \citet{2008ApJ...680.1149B} who found a value of 1.40$\times$10$^{51}$, a result of using different models. Our estimate of surface brightness is comparable to values found for galactic remnants in rarified environments, such as Tycho's SNR  ($\Sigma$ = $1.32\times 10^{-19}$~W m$^{-2}$~Hz$^{-1}$~sr$^{-1}$, $D$ = 9.3~pc) and Kepler's SNR ($\Sigma$ = $3.18\times 10^{-19}$~W m$^{-2}$~Hz$^{-1}$~sr$^{-1}$, $D$ = 5.2~pc) \citep{2013ApJS..204....4P}.

\section{Conclusion}

We have used observations taken by the ATCA to carry out a detailed radio continuum study on \SNR. With a size of only {\it D} $\cong$ 8 $\times$ 7~pc, \SNR\ is one of the smallest remnants currently known in the LMC.We find a relatively steep spectrum of ($\alpha=-0.73\pm$0.02) and relatively strong magnetic field of 168$\mu$G, which is characteristic of a young remnant (e.g. \citealt{2013MNRAS.433.1271J}). Its small size also sets this SNR apart from typical \sigmaD\ values of SNRs at  $\Sigma$~=~$1.1\times 10^{-19}$~W m$^{-2}$~Hz$^{-1}$~sr$^{-1}$, $D$ = 7.35~pc, though, still in close proximity to another Balmer dominated LMC SNR, SNR J0519--6902. This SNR shares the same radially orientated polarisation as other young Type Ia remnants, with  a mean fractional polarisation level of  $P=$~(26~$\pm$~13)\%.\\

{\bf \noindent ACKNOWLEDGEMENTS}\\

\noindent The Australia Telescope Compact Array is part of the Australia Telescope which is funded by the Commonwealth of Australia for operation as a National Facility managed by CSIRO. This research is supported by the Ministry of Education and Science of the Republic of Serbia through project No. 176005.

\bibliographystyle{mn2e}
\bibliography{0509-6731_refs}

\label{lastpage}

\end{document}